# MoOCl$_2$ as a Hyperbolic Planar Platform for Nanooptics at Telecom Frequencies.

Haozhe Tong[1,2], Clara Clemente-Marcuello[1,3], Kirill V. Voronin[1,] Pablo Alonso-González[3,4,†], Alexey Y. Nikitin[1,5,†]

[1] Donostia International Physics Center (DIPC), Donostia–San Sebastián 20018, Spain.
[2] Universidad del País Vasco/Euskal Herriko Unibertsitatea (UPV/EHU), 48940, Leioa, Spain.
[3] Department of Physics, University of Oviedo, Oviedo 33006, Spain.
[4] Center of Research on Nanomaterials and Nanotechnology, CINN (CSIC–Universidad de Oviedo), El Entrego 33940, Spain.
[5] IKERBASQUE, Basque Foundation for Science, Bilbao 48013, Spain.

[†] Corresponding authors. Email: **pabloalonso@uniovi.es**; **alexey@dipc.org**

**ABSTRACT.**
On-chip optoelectronics is fundamental to modern telecommunication, yet the diffraction limit of light remains a major obstacle to the extreme miniaturization of photonic integrated circuits (PICs). Hyperbolic polaritons (HPs) -hybrid light-matter excitations in materials with opposite-signed dielectric permittivity tensor components- offer a solution through their ability to support deep sub-wavelength confinement and unique optical phenomena such as canalization and negative refraction. To date, however, the most widely studied hyperbolic van der Waals (vdW) crystals, including hBN and $\alpha$-MoO$_3$, operate mainly in the mid-infrared, leaving the telecommunication bands (1260–1675 nm) largely uncovered. Here, we predict HPs operating directly in the telecommunication window in the vdW crystal molybdenum oxychloride (MoOCl$_2$). Building on recent evidence that MoOCl$_2$ can support plasmon polaritons in the visible, we theoretically investigate its optical response at telecom wavelengths and identify the conditions under which strongly confined, canalized HPs modes emerge. Beyond establishing a telecom platform, we outline device-level opportunities enabled by these modes, including diffraction-free waveguides based on canalization, tunable polaritonic crystals, and high-efficiency spontaneous emission-enhancement platforms. These paradigms cover the essential pillars of on-chip information processing: emission, propagation, modulation and detection. Our results establish MoOCl$_2$ as a potentially transformative material that bridges physics of hyperbolic PPs with potential practical implementations, opening avenues for ultra-compact, high-density, and low-power photonic components.

**Introduction.**

Optoelectronics, the cornerstone of modern telecommunication, relies on the efficient transmission and processing of optical signals within the 1260–1675 nm spectral window[1]. As the demand for high-density photonic integrated circuits grows, achieving extreme light confinement at the nanoscale has become a critical challenge[2]. Traditional silicon photonics is fundamentally constrained by the diffraction limit of light, necessitating the exploration of polaritons- hybrid light-matter excitations that combine the properties of photons with material oscillations such as phonons or plasmons[3,4]. VdW crystals have emerged as a premier platform for polaritons due to their atomic thickness with extreme light confinement[5,6]. In some anisotropic vdW materials, such as hBN[7], α-MoO$_3$[8] or LiV$_2$O$_5$[9] among others, the components of the permittivity tensor Re($\hat{\epsilon}$) exhibit opposite signs. This results in hyperbolic-shaped isofrequency contours (IFCs) - slices of the dispersion surface by a plane of a constant frequency, supporting HPs[10]. When the real part of the dielectric permittivity tensor elements along different crystal axes has opposite signs, the material is called hyperbolic because the IFC takes a hyperbolic shape. The latter, in its turn, means that the polariton electromagnetic field distribution has a hyperbolic-shaped wavefront[11]. Unlike traditional modes, HPs possess extremely large momenta and show unique phenomena such as negative refraction[12], canalization[13], and deeply-subwavelength focusing[14]. Despite their promise, widely studied hyperbolic vdW materials like hBN and α-MoO3 are limited to their use in mid-IR frequency range, hindering their application in technologically relevant telecommunication bands[6]. While hyperbolic metamaterials and metasurfaces can be engineered to operate at telecom frequencies[15,16], they are often plagued by significant losses and the complexities of advanced nanofabrication. Consequently, natural hyperbolic materials that support intrinsic HPs directly at telecommunication wavelengths can complement existing photonic circuits at telecom.

MoOCl$_2$ has recently been identified as a natural anisotropic vdW crystal with pronounced hyperbolicity between 0.3 eV and 2.5 eV frequencies, thus including the telecom regime[17,18]. In this work, we comprehensively investigate the optical response of thin MoOCl$_2$ layers and propose a suite of functional polaritonic architectures at telecom frequencies. We begin by characterizing previously reported dielectric permittivity tensor of MoOCl$_2$ crystal[18], as well as dispersion of PP and their field distributions in MoOCl$_2$ layers on different substrates. We then explore waveguiding paradigms, contrasting MoOCl$_2$ ribbons (potentially doable by nanostructuring of MoOCl$_2$ layers) with unpatterned MoOCl$_2$ layers on metallic substrates that support diffractionless canalized PPs. Furthermore, we mimic PCs for engineering PP band structures. The latter can be potentially controlled via rotationally-tunable architectures. Finally, we quantify light-matter interactions by calculating the Purcell enhancement of emitters near MoOCl$_2$ films, benchmarking its performance against graphene and gold.

**Results and Discussion.**

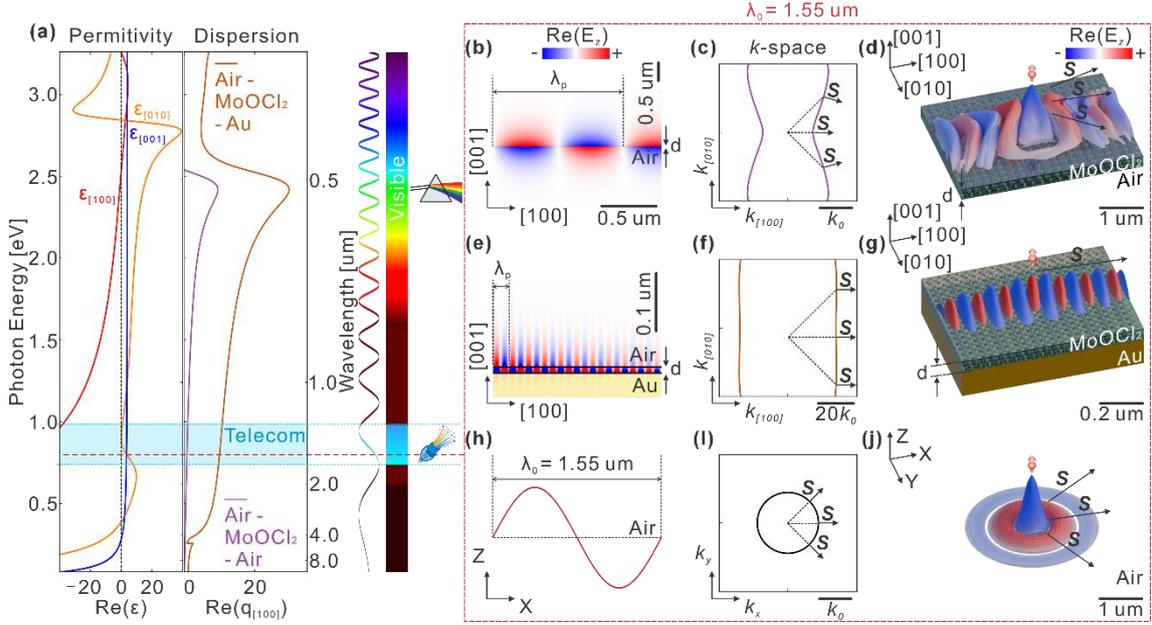

**Figure 1. Dielectric response of MoOCl$_2$ and properties of PPs supported by its thin layers. (a)** Real parts of the diagonal permittivity components, Re($\hat{\epsilon}$), (left panel) and normalized polariton momentum, Re($q_{[100]}$), (right panel) as a function of photon energy. The blue highlighted region identifies the telecommunication wavelength band (1260–1675 nm). **(b-d)** Polaritonic properties of a 10-nm-thick freestanding MoOCl$_2$ flake: **(b)** vertical electric field, Re($E_z$), distribution in $x$-$z$ plane, **(c)** IFC in $\boldsymbol{k}$-space, and **(d)** in-plane field distribution excited by a vertical point dipole 5 nm above the surface of MoOCl$_2$. **(e-g)** Polaritonic properties of a 10-nm-thick freestanding MoOCl$_2$ flake on an Au substrate: **(e)** Re($E_z$) distribution in $x$-$z$ plane, **(f)** IFC in the canalization regime with unidirectional Poynting vectors $\boldsymbol{S}$ (black arrows), and **(g)** collimated, non-diverging beam propagation. **(h-j)** properties of free-space light: **(h)** field oscillations showing the free-space wavelength, $\lambda_0$ = 1550 nm. **(i-j)** Circular IFC and isotropic field distribution provided for comparison.

In order to illustrate the propagation properties of PPs in MoOCl$_2$ layers, we first analyze their dispersion characteristics. These properties are dictated by the dielectric permittivity tensor of MoOCl$_2$, $\hat{\epsilon}$ = diag($\epsilon_{[100]}, \epsilon_{[010]}, \epsilon_{[001]}$), which is shown in the left panel of Fig.1a (adapted from Venturi et al.[18]). To facilitate our analysis, we define the Cartesian coordinates $x,y$, and $z$ as the [100], [010] and [001] crystal axes, respectively. Several components of the $\hat{\epsilon}$ tensor show strongly non-monotonic behavior in the energy range of 0.1-3.5 eV. This stems from the presence of intra-band and inter-band plasmons in MoOCl$_2$, whose frequencies appear in the Drude-Lorentz model $\varepsilon_\alpha(\omega)$ = $\varepsilon_\infty^\alpha - \frac{(\omega_p^\alpha)^2}{\omega(\omega+i\gamma_D^\alpha\omega)} + \sum_{n=1} \frac{f_n^\alpha(\omega_n^\alpha)^2}{(\omega_n^\alpha)^2-\omega^2-i\gamma_n^\alpha\omega}$, where $\alpha \in \{[100], [010], [001]\}$ denotes the crystal axis, $\omega$ is the frequency, $\varepsilon_\infty^\alpha$ is the high-frequency permittivity, while $\omega_p^\alpha$ and $\gamma_D^\alpha$ represent the plasma frequency of intraband free carrier oscillations and the corresponding Drude scattering rate, respectively[18]. The summation term represents interband electronic transitions modeled by Lorentz oscillators, where $f_n^\alpha$, $\omega_n^\alpha$ and $\gamma_n^\alpha$ are the oscillator strengths, resonance frequencies and scattering rates of the $n$-th Lorentz oscillator, respectively. The plasma frequency $\omega_p^\alpha$ (where the real part of the dielectric permittivity typically crosses zero) along the [010] and [001] axes fall within the mid-IR frequency range, whereas for the [100] direction it resides in the visible range. The opposite signs of different components of Re($\hat{\epsilon}$) define the so-called hyperbolic region [7,8]. In particular, at telecommunication frequencies (highlighted in blue in Fig.1a), the real part of $\varepsilon_{[100]}$ (red curve) is negative, whereas the real parts of $\varepsilon_{[010]}$ (yellow curve) and $\varepsilon_{[001]}$ (blue curve) remain positive. Thus, according to the dielectric permittivity model and fitted parameters, the hyperbolic region encompasses the entire telecommunication range. To provide a comprehensive spatial and directional analysis, all subsequent results in Figs. 1b–j are shown at the telecommunication C-band center wavelength ($\lambda_0$=1550 nm)[19], corresponding to the minimum loss window of silica fibers (marked by the red dashed line in Fig.1a). To characterize the dispersion of PPs in a $MoOCl_2$ layer, we calculate the normalized momentum $q = k_P/k_0$ using the following analytical expression[20,21], valid in the limit of large polariton wavevectors:

$$q = \frac{\rho}{k_0 d}\left[arctan\left(\frac{\rho \varepsilon_1}{\varepsilon_z}\right) + arctan\left(\frac{\rho \varepsilon_3}{\varepsilon_z}\right) + \pi l\right]$$

Where $k_P$ and $k_0$ is the wavevector of PPs and light in free space, respectively; $\rho = i\sqrt{\frac{\epsilon_z}{\epsilon_x cos^2\psi + \epsilon_y sin^2\psi}}$, $\psi$ is the polar angle between $\mathbf{k_P}$ and the $x$-axis; $\epsilon_1$ and $\epsilon_3$ are the dielectric permittivities of the semi-infinite substrate and superstrate, respectively; $d$ is the thickness of the $MoOCl_2$ layer; and $l$ is the mode index ($l = 0,1,...$). This equation determines the momenta of an infinite set of PP modes, $M_l$, supported by the layer. As indicated by the expression, the PP momentum increases with thickness reduction and is highly sensitive to the surrounding dielectric environment, as defined by the arctan functions terms containing $\epsilon_1, \epsilon_3$ in their arguments. Notice that the propagation length of the PP $M_l$ modes decreases with $l$, consequently, the lowest-momentum (fundamental) mode, $M_0$, is typically dominant in hyperbolic biaxial layers on transparent substrates[21]. For the sake of simplicity, but without loss of generality, we consider a freestanding 10-nm-thick $MoOCl_2$ (Air-$MoOCl_2$-Air, $\epsilon_1 = \epsilon_3 = 1$), assuming low refracting index substrates. In the right panel of Fig.1a the dispersion relation (Re($q$) as a function of frequency) for the fundamental PP $M_0$ is close to 1 (i.e. $k_P \approx k_0$), indicating that the PPs are virtually indistinguishable from the light in free space and are weakly confined to the $MoOCl_2$ layer. This weak confinement is substantiated by the spatial electric field distribution, Re($E_z$), in the $x$-$z$ plane presented in Fig.1b. Consistent with the dispersion relation, the oscillation period of PPs, $\lambda_p$ (the distance between fringes of opposite polarities represented by red and blue colors), nearly matches the wavelength of light in free space, $\lambda_0$, as illustrated in Fig.1h. To evaluate the propagation of PPs in different in-plane directions, the calculated isofrequency contour IFC, is presented in Fig.1c. The IFC exhibits a hyperbolic-like shape, although not strongly deviating from the circle IFC of free-space light (a cross-section of the light cone, Fig.1l). The Poynting vector $\mathbf{S} = \mathbf{E} \times \mathbf{H}$ is perpendicular to the IFC and is generally misaligned with the momentum $\mathbf{k}$ (except along [100] direction)[10]. This intrinsic anisotropy of PPs is further evidenced by the simulated electric field Re($E_z$) distribution generated by a vertical point dipole source above the $MoOCl_2$ layer (Fig.1d), where the wavefront morphology aligns with the hyperbolic-like IFC. Consequently, while the freestanding $MoOCl_2$ layer supports anisotropic PPs, they exhibit low field confinement in this configuration. In the context of existing technologies, the confinement factor ($q = k_p/k_0$) of PPs in freestanding $MoOCl_2$ is notably lower than that of conventional Silicon-on-Insulator (SOI) waveguides, which typically exhibit an effective refractive index ($n_{eff}$, equivalent to $q$) around 2 for standard 220-nm-thick strip configurations at 1550 nm[22]. However, the intrinsic hyperbolic anisotropy of MoOCl$_2$ offers distinctive physical advantages unavailable to isotropic photonic platforms. Specifically, the open topology of the hyperbolic IFCs supports modes with arbitrarily large momentum in certain directions, providing a path to overcome the diffraction limit. Furthermore, this in-plane anisotropy introduces unique rotational tunability for directional light steering and polarization routing, enabling precise manipulation of polariton propagation via crystal orientation[13]. More importantly, this baseline configuration sets the stage for achieving extreme sub-wavelength confinement and "canalized" propagation through substrate engineering[23,24].

Interestingly, when MoOCl$_2$ is placed on a substrate with a large absolute dielectric permittivity, $|\epsilon_3| \gg 1$ (including metals with large negative permittivities), the M$_0$ mode is not supported. Instead, M$_1$ mode becomes the dominant fundamental mode[25]. The momentum of this mode reaches significantly larger values, as shown by its calculated dispersion relation in Fig.1b (gold curve) for MoOCl$_2$ on a gold substrate (Air-MoOCl$_2$-Au, $\epsilon_1 = 1$, $\epsilon_3 = \epsilon_{Au}$). This larger momentum implies a smaller wavelength, $\lambda_p$, and consequently stronger field confinement, as illustrated in Fig.1e. Another remarkable feature of PPs in a MoOCl$_2$ layer on a metallic substrate is that their IFCs (shown in Fig.1f) consist of two nearly parallel lines positioned far away from the IFC of light in free space (black circle, representing the cross-section of the light cone in Fig.1l). This separation signifies the transition to a high-momentum regime with extreme field localization. Impressively, the Poynting vector $S$ (black arrows in Fig.1f,g) remains unidirectional across all the wavevectors on the IFC, signifying the so-called canalization regime[13,23]. This behavior is clearly visible in the Re($E_z$) field distribution in Fig.1g, where a vertical dipole source excites a highly collimated, non-diverging beam along the [100] axis. This canalization effect allows the polariton to propagate as a "diffractionless" ray, a property that can be potentially important for the routing of optical information in high-density telecommunication architectures.

These results demonstrate that MoOCl$_2$ hosts highly anisotropic PPs characterized by deep sub-wavelength electromagnetic field confinement. By employing precise substrate engineering, we observe the transition to a canalized propagation regime where the polariton flow is strictly guided along a preferred crystal axis with minimal spatial divergence, representing a distinct propagation mode that differs from traditional diffraction-limited waveguiding.

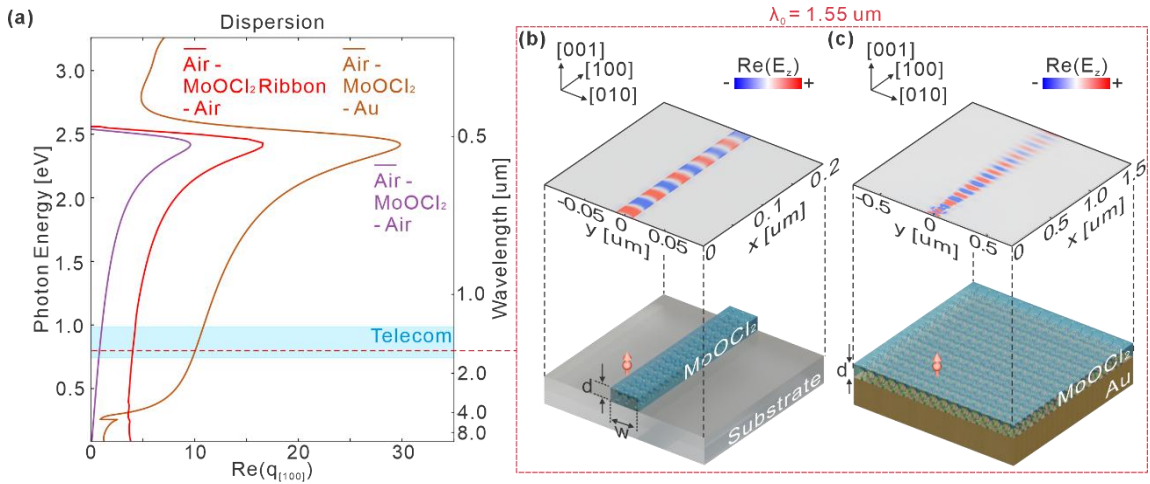

**Figure 2. Nanoscale waveguiding with thin MoOCl$_2$ layers. (a)** Comparison of the normalized momentum, $q$, along the [100] crystal direction as a function of the photon energy for freestanding MoOCl$_2$ (blue curve), a 20-nm-wide nanoribbon (red curve), and a MoOCl$_2$ slab on the gold substrate (gold curve). **(b)** 3D schematic (lower) and simulated Re($E_z$) field (upper) for a MoOCl$_2$ nanoribbon ($d$ = 10 nm, $w$ = 20 nm) on a dielectric substrate, demonstrating geometric confinement. **(c)** 3D schematic (lower) and simulated Re($E_z$) field (upper) for unpatterned MoOCl$_2$ layer on a gold substrate utilizing the canalization effect, showing the collimated field distribution excited by a dipole source at $\lambda_0$ = 1550 nm.

Leveraging the fundamental polaritonic properties in MoOCl$_2$ layers, specifically its extreme field confinement and directional energy flow, we propose two distinct paradigms for nanoscale waveguiding at telecommunication frequencies. Consistent with our previous analysis, both scenarios are evaluated at the C-band center wavelength ($\lambda_0$ = 1550 nm). First, we theoretically investigate the feasibility of structuring MoOCl$_2$ layers on dielectric substrates (e.g., CaF$_2$, SiO$_2$, Si or similar) into conventional strip-like waveguides. Fig.2b illustrates an example of a MoOCl$_2$ nanoribbon (thickness $d$ = 10 nm, width $w$ = 20 nm) embedded in a dielectric environment ($\epsilon_1$ = $\epsilon_3$ = 1). The normalized momentum of the fundamental ribbon mode is represented by the red curve in Fig.2a. The momentum of the ribbon mode is significantly larger than that of a freestanding 10nm-thick infinite MoOCl$_2$ slab (purple curve), confirming that nanofabrication-induced lateral boundaries impose additional geometric light confinement[26]. The upper panel of Fig.2b displays the simulated electric field distribution, Re($E_z$), within the $x$-$y$ plane 0.1 nm above the MoOCl$_2$ surface, illustrating the field profile excited by a vertical electric dipole source positioned 5 nm above the surface. We clearly observe spatially confined oscillations of the propagating fundamental PP mode along the ribbon, demonstrating effective sub-wavelength waveguiding mediated by geometric patterning.

In the second scenario, we theoretically examine a MoOCl$_2$ layer integrated onto a metallic substrate. In stark contrast to patterned ribbons, the MoOCl$_2$ layer on metal can remain unpatterned, so that its lateral dimensions are effectively infinite relative to the polariton confinement length. Crucially, by leveraging the canalization effect induced through substrate engineering, we can realize unidirectional waveguiding using a native, unpatterned MoOCl$_2$ slab on a gold (Au) substrate (Fig.2c, lower panel). The calculated normalized momentum $q$ for a 10 nm thick MoOCl$_2$ slab on Au is shown as the gold curve in Fig.2a, which reveals significantly enhanced wavelength compression compared to dielectric-supported configurations. The corresponding simulated Re($E_z$) field distribution within the $x$-$y$ plane 0.1 nm above the MoOCl$_2$, excited by a dipole source positioned 5 nm above the surface, is presented in the upper panel of Fig.2c. Consequently, MoOCl$_2$ emerges as a potential superior waveguiding platform that achieves extreme field localization in its native state, bypassing the need for complex nanofabrication.

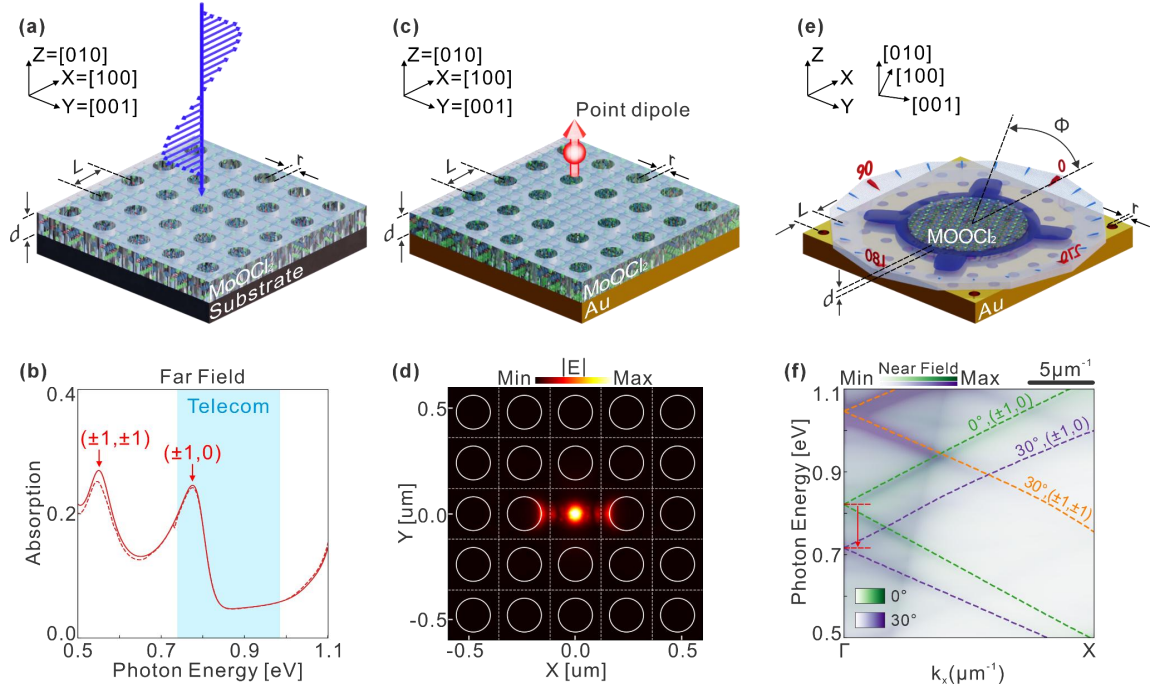

**Figure 3. MoOCl$_2$ polaritonic crystals. (a-b)** Schematic of a hole array made in freestanding MoOCl$_2$ PC (array period $L$ = 1.1 µm) and its absorption spectrum under illumination by a normally-incident plane wave polarized along the x-axis, revealing resonant peaks for (±1,0) and (±1,±1) orders. The solid and dashed lines represent the simulation and the analytical results, respectively. **(c-d)** schematic of MoOCl$_2$ PC cavity on an Au substrate and the electric field absolute value, |***E***|, distribution, showing deep sub-wavelength energy localization. **(e-f)** Twisted MoOCl$_2$/patterned-Au heterostructure: **(e)** schematics showing an MoOCl$_2$ layer twisted by angle $\Phi$ with respect to the hole array in Au and **(f)** calculated band structure of PPs. The red arrow indicates a spectral shift of the PP bands from 6530 cm$^{-1}$ (green, $\Phi$=0°) to 5710 cm$^{-1}$ (purple, $\Phi$=30°).

Furthermore, to expand the versatility of MoOCl$_2$ for signal control, we consider MoOCl$_2$ PCs - periodically-structured materials supporting polaritons. Similarly to conventional photonic crystals[27–29], PCs are engineered such that their lattice constant (period), $L$, is commensurate with the PP wavelength $\lambda_p$, enabling matching between incident in-plane light momentum, $k_t$, and high-momentum of PP modes, dictated by the IFCs. We first theoretically consider a freestanding MoOCl$_2$ PC formed by patterning a periodic square hole array (period $L$ = 1.1 µm and hole radius $r$ = 0.4 µm) in a 10 nm thick MoOCl$_2$ slab (Fig.3a). The lattice is defined by two orthogonal lattice vectors, $\boldsymbol{L_1}$ and $\boldsymbol{L_2}$, along $x$ and $y$ directions respectively ($|\boldsymbol{L_1}|=|\boldsymbol{L_2}|=L$). In this configuration, the M$_0$ mode's momentum, $k_P$, remains relatively close to the free-space wavevector $k_0$, facilitating efficient coupling through the periodic lattice. Fig.3b displays the calculated absorption spectrum for normal $x$-polarized incidence ($k_{in} = k_0$, the in-plane momentum of the incidence light $k_{t,in}$ = 0). The full-wave simulations (solid line) and analytical approximation (dashed line, see Methods) show an excellent agreement, revealing two distinct absorption peaks. The periodicity of the square PC defines discrete reciprocal lattice vectors $\boldsymbol{G}(n_1,n_2) = \frac{2\pi}{L}(n_1,n_2)^{\text{T}}$, where $n_1$ and $n_2$ are integers representing the diffraction orders. These vectors compensate for the momentum mismatch between incident photons and high-momentum polariton modes. Under illumination, polaritons are resonantly excited when the momentum-matching condition $\boldsymbol{k_{t,in}} + \boldsymbol{G}(n_1, n_2) = \boldsymbol{k_p}$ is met[13]. The two observed peaks in the spectrum correspond to ($\pm 1, 0$) and ($\pm 1, \pm 1$) resonance orders, respectively.

While the $M_0$ PP mode in freestanding $MoOCl_2$ facilitates coupling due to its low momentum ($k_P \approx k_0$), its large polariton wavelength $\lambda_p$ limits the spatial compactness of freestanding PCs. To achieve appreciable device miniaturization, we leverage the significant wavelength compression provided by a thin $MoOCl_2$ layer on a gold substrate[25]. The architecture illustrated in Fig.3c consists of a $MoOCl_2$ hole-array PC on Au substrate ($L$ = 240 nm, $r$ = 80 nm). As a proof-of-concept for high-density integration, we demonstrate a PC defect cavity, consisting of a single missing hole in the PC[30]. Fig.3d illustrates the spatial distribution of the absolute electric field, $|\boldsymbol{E}|$, at the $MoOCl_2$ surface, excited by a vertical dipole positioned 50 nm above the defect. We observe that the field is strongly confined within the defect area, a phenomenon associated with the formation of a polaritonic bandgap in the PC[31]. Specifically, when the Bragg condition is satisfied, $k_t \approx |\boldsymbol{G}(n_1, n_2)|$, counter-propagating PP modes connected by a reciprocal lattice vector become nearly degenerate and undergo a coupling via Bragg scattering. This interaction induces a forbidden frequency range, a polaritonic bandgap, at the Brillouin-zone boundaries. Because propagation is prohibited for frequencies within this gap, the electromagnetic field is "trapped" at the defect volume, forming a cavity mode with deep sub-wavelength localization.

The inherent anisotropy of PPs in $MoOCl_2$ introduces an additional degree of freedom for tuning PCs: the twist angle between the crystal axes and the lattice vectors, according to a recently introduced concept of twisted PCs[25]. However, directly patterning $MoOCl_2$ flakes lacks post-fabrication tunability. To circumvent this, we suggest using a periodically-engineered substrate rather than the $MoOCl_2$ flake itself [25]. Fig.3e illustrates this structure, where a $MoOCl_2$ flake is placed onto a square hole-array in a gold substrate ($L$ = 160 nm, $r$ = 15 nm). The $x$ and $y$ axes are aligned with the lattice vectors, while the twist angle $\Phi$ is defined between the $MoOCl_2$ [100] axis and the $x$-axis. Fig.3f illustrates the calculated band structure for two values of $\Phi$. The dashed lines represent the empty lattice dispersion, which depicts the polariton modes folded into the first Brillouin zone in the absence of lattice-induced perturbation. In the band structure, we observe a significant spectral shift between $\Phi = 0°$ (green) and $\Phi = 30°$ (purple). Specifically, the red arrow highlights that the bands near the $\Gamma$ point shift from 6530 cm$^{-1}$ to 5710 cm$^{-1}$ as $\Phi$ increases. This pronounced frequency shift originates from the alignment between the anisotropic IFC and the substrate's reciprocal lattice. Indeed, at $\Phi = 0°$, the IFC matches the Bragg condition for the ($\pm 1, 0$) lattice vectors at 6530 cm$^{-1}$. Upon rotating the flake to $\Phi = 30°$, the anisotropic IFC realigns, meeting the momentum-matching requirement at a lower frequency of 5710 cm$^{-1}$. The shift of the PP bands with $\Phi$ demonstrates that PCs based on $MoOCl_2$ layers on nanostructured metallic substrate provide a highly (and potentially actively-) tunable platform for polariton bandgap engineering. In summary, the implementation of $MoOCl_2$ PCs and defect cavities demonstrates a transition toward functional nanophotonic components. In contrast to conventional silicon-based PCs limited by the diffraction limit, or vdW platforms like hBN and α-$MoO_3$ that are restricted to the mid-infrared[32,33], $MoOCl_2$ enables extreme sub-wavelength confinement across the 1260–1675 nm telecom window. Our results show that exploiting the material's intrinsic anisotropy through rotational heterostructure tuning allows for a dynamic, post-fabrication spectral shift of the polaritonic bandgap. This distinguishes MoOCl$_2$ from traditional static PC technologies, providing a reconfigurable degree of freedom for integrated telecom circuits that are typically absent in standard semiconductor architectures.

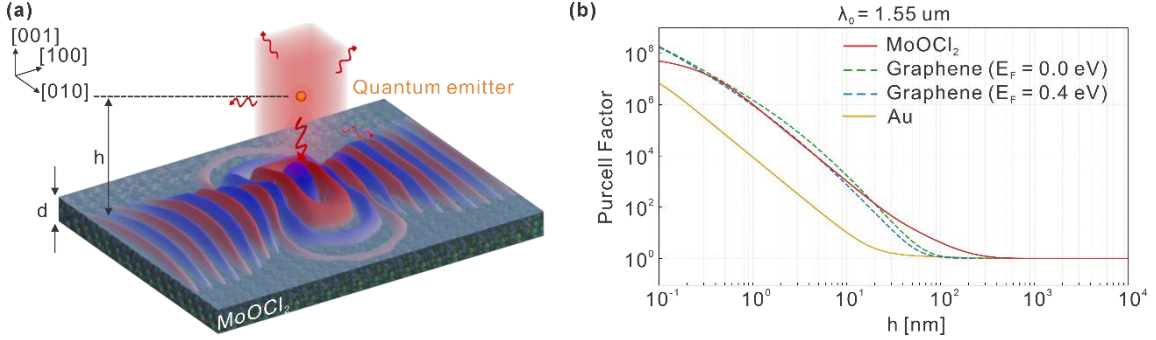

**Figure 4. Spontaneous emission enhancement at telecommunication wavelengths. (a)** Schematics of a quantum emitter at height $h$ above a 10-nm-thick MoOCl$_2$ slab. The calculated field distribution is shown by the 3D surface. **(b)** Calculated Purcell factor as a function of $h$ at $\lambda_0$ = 1550 nm. MoOCl$_2$ layer (red solid line) is benchmarked against graphene at different Fermi levels ($E_F$ = 0, 0.4 eV, dashed lines) and bulk Au (yellow solid line). MoOCl$_2$ provides superior enhancement for $h$ > 20 nm due to increased local density of states.

To evaluate the potential of MoOCl$_2$ for quantum optics applications at telecom wavelengths, we investigate its performance for spontaneous emission enhancement. The studied configuration is illustrated in Fig.4a, where a vertical dipole source, mimicking a quantum emitter (which can be, for example a quantum dot[34]) is positioned at a distance $h$ above a MoOCl$_2$ slab of thickness 10 nm. The emitter's near-field and triggers the excitation of high-momentum hyperbolic polaritons in the MoOCl$_2$ layer, providing an additional fast radiative decay channel for the emitter[35–37]. To quantify this effect, we calculate the Purcell enhancement factor, $F_P = \Gamma/\Gamma_0$, where $\Gamma$ is the decay rate of the emitter near the surface, $\Gamma_0$ represents its decay rate of emitters in free space. We calculate the Purcell Factor for biaxial material using the following formula [35]

$$F_P = \frac{3}{2} Re[\int_0^\infty dq \frac{q^3}{q_z}(q - r_p(q)e^{2ik_0 q_z h})]$$

where $q_z = \sqrt{1 - q^2}$ is the momentum in the direction perpendicular to the layer, with Im($q_z$) > 0, and $r_p(q) = -\alpha q_z/(\alpha q_z + 1)$ is the reflection coefficient of the MoOCl$_2$ layer for p-polarization, with $\alpha = 2\pi\sigma_{eff}/c$ being the effective normalized conductivity, $\sigma_{eff} = \frac{\omega d\epsilon}{4\pi}$ is the effective 2D conductivity of the slab[38]. In Fig.4b, we present the calculated Purcell factor for the freestanding MoOCl$_2$ slab compared to a monolayer graphene and a bulk gold at a wavelength of 1550 nm. Our results reveal a distinct performance according to the emitter-surface separation, $h$. In the extreme near-field regime ($h$ < 20 nm), the Purcell enhancement of MoOCl$_2$ is comparable to that of graphene. However, as the distance increases beyond $h$ = 20 nm, MoOCl$_2$ significantly outperforms graphene, maintaining a higher Purcell factor over a broader range of separation. This sustained enhancement originates from the increased electromagnetic local density of states (LDOS) near the MoOCl$_2$ surface because of the confined PPs in MoOCl$_2$[39,40]. These results highlight MoOCl$_2$ as a superior and more versatile platform for enhancing light-matter interactions in next-generation integrated photonic and quantum telecommunication systems[41,42].

**CONCLUSION AND OUTLOOK.**

Through a comprehensive approach combining analytical modeling and numerical simulations, we investigated the Molybdenum oxychloride's unique optical response and hyperbolic polariton characteristics, positioning it as a transformative material for the potential next generation of ultra-compact, high-density, and low-power photonic integrated circuits. Specifically, we proposed two distinct paradigms for nanoscale waveguiding: effective sub-wavelength propagation in geometrically patterned $MoOCl_2$ nanoribbons on dielectric substrates, and, more notably, "diffraction-free" canalization in an unpatterned $MoOCl_2$ layer on a metallic substrate, a unidirectional propagation mode along a preferred crystal axis that bypasses the need for complex nanofabrication. Furthermore, we introduced twist-tunable $MoOCl_2$ PCs and defect cavities, which enable deep sub-wavelength energy localization and signal control. They can provide a dynamic spectral shift of the polaritonic bandgap, thus offering a reconfigurable modulation capability typically absent in traditional static semiconductor architectures. Finally, our calculations confirmed $MoOCl_2$'s potential for quantum optics applications, showing superior Purcell enhancement for quantum emitters at emitter-surface separations greater than 20 nm when compared to graphene and bulk gold, a performance stemming from the enhanced LDOS near the $MoOCl_2$ surface. Collectively, our results establish $MoOCl_2$ as a material platform that effectively bridges the fundamental physics of hyperbolic polaritons within practical telecommunication frequency bands (1260–1675 nm). Future efforts should be directed toward further engineering design and the crucial experimental verification of the proposed components, including the waveguides, polaritonic crystals, and emission enhancement platforms, to accelerate the adoption of $MoOCl_2$ in compact, telecom-band optical circuits.

**METHODS**

**Dielectric Permittivity Modeling.** The biaxial dielectric response of $MoOCl_2$ was modeled using a diagonal permittivity tensor $\hat{\epsilon}$ = diag($\epsilon_{[100]}, \epsilon_{[010]}, \epsilon_{[001]}$), aligned with the [100], [010], and [001] crystal directions. Each component was modeled using a combined Drude-Lorentz framework[18]: $\varepsilon_\alpha(\omega) = \varepsilon_\infty^\alpha - \frac{(\omega_p^\alpha)^2}{\omega(\omega + i\gamma_D^\alpha \omega)} + \sum_{n=1} \frac{f_n^\alpha(\omega_p^\alpha)^2}{(\omega_n^\alpha)^2 - \omega^2 - i\gamma_n^\alpha \omega}$, where $\alpha \in \{[100], [010], [001]\}$. The specific parameters for each axis are as follows: $\varepsilon_\infty^{[100]}$ = 2.74, $\omega_p^{[100]}$ = 6.419 eV, $\gamma_D^{[100]}$ = 0.023 eV, $f_n^{[100]}$ = 0.42 eV, $\omega_n^{[100]}$ = 3.28 eV, $\gamma_n^{[100]}$ = 0.46 eV; $\varepsilon_\infty^{[010]}$ = 3.69, $\omega_p^{[010]}$ = 1.361 eV, $\gamma_D^{[010]}$ = 0.016 eV, $f_n^{[010]}$ = 14 eV, $\omega_n^{[010]}$ = 2.84 eV, $\gamma_n^{[010]}$ = 0.13 eV; $\varepsilon_\infty^{[001]}$ = 3.96, $\omega_p^{[001]}$ = 0.562 eV, $\gamma_D^{[001]}$ = 0.025 eV. The dielectric permittivity of the gold (Au) was described using a standard Drude model with parameters adapted from Rodrigo et al.[43].

**Numerical Simulations.** All full-wave electromagnetic simulations were performed using COMSOL Multiphysics based on the finite-element method (FEM) in the frequency domain. For the near-field distributions presented in Fig.1, we modeled a $MoOCl_2$ layer with a thickness $d$ = 10 nm, excited by a vertical electric point dipole positioned 5 nm above the crystal surface; the real part of the vertical electric field, Re($E_z$), was recorded on a plane 1 nm above the flake. For the waveguiding structures in Fig.2, the same thickness ($d$ = 10 nm) was applied to both the MoOCl$_2$ slab and the nanoribbons (width $w$ = 20 nm), maintaining the dipole excitation and field monitoring heights at 5 nm and 1 nm, respectively. Regarding the PCs shown in Fig.3, the absorption spectra of the freestanding configuration were calculated using an infinite MoOCl$_2$ layer ($d$ = 10 nm) with a hole array defined by a period $L$ = 1.1 μm and radius $r$ = 0.4 μm under $x$-polarized normal incidence. Finally, the localized field distribution $|E|$ in the PC defect cavity was simulated by modeling a hole array ($L$ = 240 nm, $r$ = 80 nm) on a gold substrate with a single vacancy at the center, where the field was extracted from the MoOCl$_2$ surface under excitation by a dipole placed 50 nm above the structure.

**Analytical Calculations.** The dispersion relations and iso-frequency contours (IFCs) of hyperbolic polaritons (HPs) in MoOCl$_2$ layers were analytically determined using the large-momentum approximation ($q = k_P/k_0 \gg 1$). For a biaxial crystal slab of thickness $d$ surrounded by media with permittivities $\epsilon_1$ and $\epsilon_3$, the normalized momentum $q$ as a function of the in-plane propagation angle $\psi$ (relative to the [100] axis) is given by $q = \frac{\rho}{k_0 d}\left[\arctan\left(\frac{\rho\varepsilon_1}{\varepsilon_z}\right) + \arctan\left(\frac{\rho\varepsilon_3}{\varepsilon_z}\right) + \pi l\right]$, where $l = 0,1,\ldots$ is the mode index, and the structural anisotropy is encapsulated in the variable $\rho = i\sqrt{\frac{\epsilon_z}{\epsilon_x cos^2\psi + \epsilon_y sin^2\psi}}$. This framework was employed to calculate the dispersion curves in Fig.1 and Fig.2 for a MoOCl$_2$ layer with $d$ = 10 nm, considering both freestanding ($\epsilon_1 = \epsilon_3 = 1$) and gold-substrate ($\epsilon_3 = \epsilon_{Au}$) configurations. For the polaritonic crystals in Fig.3, the analytical band structures and absorption spectra were calculated following the transfer-matrix and scattering-matrix methods adapted from Capote-Robayna et. al.[25,33]. The spontaneous emission enhancement in Fig.4 was evaluated via the Purcell factor, $F_P = \frac{3}{2} Re[\int_0^\infty dq \frac{q^3}{q_z}(q - r_p(q)e^{2ik_0 q_z h})]$, where $q_z = \sqrt{1-q^2}$ is the normalized out-of-plane momentum (Im($q_z$) > 0), $h$ is the emitter-surface distance, and $r_p(q) = -\alpha q_z/(\alpha q_z + 1)$ is the p-polarized reflection coefficient. Here, $\alpha = 2\pi\sigma_{eff}/c$ represents the normalized effective conductivity, with the slab treated as an effective 2D medium where $\sigma_{eff} = \frac{\omega d \epsilon}{4\pi}$[38].

## ASSOCIATED CONTENT
**Notes**
The authors declare no competing financial interest.

## ACKNOWLEDGMENTS
The study was funded by the Department of Science, Universities and Innovation of the Basque Government (grant PIBA-2023-1-0007) and the IKUR Strategy; by the Spanish Ministry of Science and Innovation (grants PID2023-147676NB-I00 and PID2022-141304NB-I00). P.A.-G. acknowledges financial support from the ERC under Consolidator grant number 101044461.